 \documentclass[10pt,twocolumn,twoside]{IEEEtran}\newcommand{\VERSION}{FINAL}

\usepackage{cite}
\usepackage{array}
\usepackage{fixltx2e}
\usepackage{color}
\usepackage{psfrag}
\usepackage{epsfig}
\usepackage{tabularx}
\usepackage{amsmath}
\usepackage{amssymb}
\usepackage{amsfonts}
\usepackage{algorithm,algorithmic}
\usepackage{pstricks, pst-node, pst-plot, pst-circ}
\usepackage{moredefs}
\usepackage{ifthen}

\def\PSNR{\mathrm{ PSNR}}

\def\punit{\, \mathrm}

\def\e{\mathrm{ e}}

\def\j{\mathrm{ j}}

\newcommand{\argmax}{\mathop{\mathrm{argmax}}}

\newcolumntype{C}[1]{>{\centering\arraybackslash}p{#1}}

\newcommand{\algorithmicinput}{\textbf{input:}}
\newcommand{\INPUT}{\item[\algorithmicinput]}
\newcommand{\algorithmicoutput}{\textbf{output:}}
\newcommand{\OUTPUT}{\item[\algorithmicoutput]}

\newcolumntype{C}[1]{>{\centering\arraybackslash}p{#1}}

\begin{document}


\title{Complex-Valued Frequency Selective Extrapolation for Fast Image and Video Signal Extrapolation}
\author{J{\"u}rgen~Seiler${}^\ast$,~\IEEEmembership{Student~Member,~IEEE}
        and~Andr{\'e}~Kaup,~\IEEEmembership{Senior~Member,~IEEE}
\thanks{ J. Seiler and A. Kaup are with the Chair of Multimedia Communications and Signal Processing, University of Erlangen-Nuremberg, Cauerstr. 7, 91058 Erlangen, Germany (e-mail: seiler@lnt.de; kaup@lnt.de, \mbox{phone: +49 9131 85 27103}, \mbox{fax: +49 9131 85 28849}). }}

\markboth{}{Seiler and Kaup: Complex-Valued Frequency Selective Extrapolation}

\maketitle


\begin{abstract} \label{abstract}
Signal extrapolation tasks arise in miscellaneous manners in the field of image and video signal processing. But, due to the widespread use of low-power and mobile devices, the computational complexity of an algorithm plays a crucial role in selecting an algorithm for a given problem. Within the scope of this contribution, we introduce the complex-valued Frequency Selective Extrapolation for fast image and video signal extrapolation. This algorithm iteratively generates a generic complex-valued model of the signal to be extrapolated as weighted superposition of Fourier basis functions. We further show that this algorithm is up to 10 times faster than the existent real-valued Frequency Selective Extrapolation that takes the real-valued nature of the input signals into account during the model generation. At the same time, the quality which is achievable by the complex-valued model generation is similar to the quality of the real-valued model generation. 
\end{abstract}

\ifthenelse{\equal{\VERSION}{DRAFT}}{{\bf EDICS:} IMD-ANAL, IMD-MDSP, IMD-CODE}{}


\section{Introduction} \label{sec:introduction}

 \IEEEPARstart{T}{he} extrapolation of signals is a very important task in image and video signal processing. The necessity for extending a signal from known areas into unknown ones arises e.\ g.\ in the area of concealing image distortions caused by transmission errors \cite{Stockhammer2005}. A similar task is inpainting \cite{Bertalmio2000}, whose aim is to remove objects or flaws from images in an invisible manner. For inpainting, where even image database based approaches \cite{Hays2007} exist, computational complexity is a less important issue as for error concealment in image an video communication. There distortions have to be removed fast and with low computational load. As presented in \cite{Olshausen1997}, sparsity based algorithms are well suited for underdetermined extrapolation problems and can be applied to image signals as these are sparse in certain transform domains according to \cite{Candes2007}. In \cite{Tropp2004}, it is shown that greedy sparse algorithms are of special interest, as these algorithms are able to robustly carry out the extrapolation. One powerful algorithm out of this group is the Selective Extrapolation from \cite{Kaup2005} and its special realization operating in the Fourier domain, the Frequency Selective Extrapolation (FSE). In the meantime, this algorithm has been adopted by several others \cite{Herraiz2008, Friebe2007} to solve extrapolation problems in their context.

Although FSE as proposed in \cite{Kaup2005} for image signal extrapolation is able to achieve a very high extrapolation quality, it has the drawback of a relatively high computational complexity which would be especially harmful for mobile devices. The high computational complexity results from the fact that the original FSE generates a real-valued model of the signal. Subsequently, we show that by omitting this constraint and generating a complex-valued model even for real-valued input signals, the complexity can be significantly reduced. As we will show at the end, the complex-valued model generation is able to be up to $10$ times faster than the original FSE at the same extrapolation quality.


\section{Real-Valued Frequency Selective Extrapolation} \label{sec:rfse} 

\begin{figure}
	\centering
	\psfrag{m}[c][c]{$m$}
	\psfrag{n}[c][c]{$n$}
	\psfrag{Loss area}[l][l]{Loss area $\mathcal{B}$}
	\psfrag{Support area}[l][l]{Support area $\mathcal{A}$}
	\ifthenelse{\equal{\VERSION}{DRAFT}}{
		\includegraphics[width=0.4\textwidth]{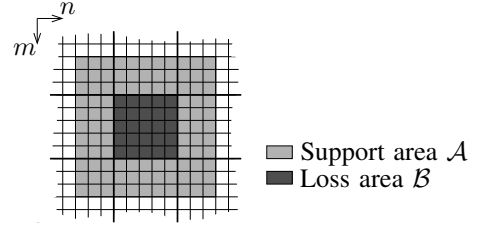}}{
		\includegraphics[width=0.25\textwidth]{graphics/extrapolation_area.eps}\vspace{-.2cm}}
	\caption{Extrapolation area $\mathcal{L}$ consisting of loss area $\mathcal{B}$ and support area $\mathcal{A}$.}
	\label{fig:extrapolation_area}
\end{figure}

Before the complex-valued Frequency Selective Extrapolation is outlined in detail, the original Frequency Selective Extrapolation which is proposed in \cite{Kaup2005} for image and video signal extrapolation is reviewed briefly. Although the algorithm can be easily extended to higher-dimensional data sets by making use of \cite{Meisinger2007}, for presentational reasons it will be regarded for two-dimensional data sets only, here. The extrapolation aims at recovering a signal $s\left[m,n\right]$ in the so called loss area $\mathcal{B}$ from known values in support area $\mathcal{A}$. Areas $\mathcal{A}$ and $\mathcal{B}$ together form extrapolation area $\mathcal{L}$ which is depicted by the coordinates $m$ and $n$ and is of size $M\times N$. Fig.\ \ref{fig:extrapolation_area} shows these areas for the example of extrapolation of a lost block from surrounding, known samples. 

The algorithm which is proposed in \cite{Kaup2005} for image and video signal extrapolation generates a real-valued model by superimposing Fourier basis functions. Hence, it is denoted by real-valued Frequency Selective Extrapolation (rFSE), subsequently. The basis functions on which the model is based emanate from the Discrete Fourier Transform (DFT) and are defined by:
\begin{equation}
\label{eq:fourier_bf}
 \varphi_{\left(k,l\right)}\left[m,n\right] = \e^{\j \frac{2\pi}{M}km}\e^{\j \frac{2\pi}{N}ln}.
\end{equation}
For the extrapolation, rFSE iteratively generates the model 
\ifthenelse{\equal{\VERSION}{DRAFT}}{
\begin{equation}
\label{eq:rFSE_model}
 g\left[m,n\right] = \frac{1}{2}\sum_{\left(k,l\right)\in\mathfrak{K}}\left(\hat{c}_{\left(k,l\right)} \varphi_{\left(k,l \right)}^\ast\left[m,n\right] + \hat{c}_{\left(k,l \right)}^\ast \varphi_{\left(k,l \right)}\left[m,n\right]\right).
\end{equation}
}{
\begin{equation}
\label{eq:rFSE_model}
 g\left[m,n\right] \hspace{-1mm}=\hspace{-1mm} \frac{1}{2}\hspace{-1.5mm}\sum_{\left(k,l\right)\in\mathfrak{K}}\hspace{-3mm}\left(\hat{c}_{\left(k,l\right)} \varphi_{\left(k,l \right)}^\ast\left[m,n\right] \hspace{-1mm}+\hspace{-1mm} \hat{c}_{\left(k,l \right)}^\ast \varphi_{\left(k,l \right)}\left[m,n\right]\right).
\end{equation}
}
as weighted superposition of basis functions. The weighting factors $\hat{c}_{\left(k,l\right)}$ and $\hat{c}_{\left(k,l \right)}^\ast$ respectively are called expansion coefficients and set $\mathfrak{K}$ holds the index tuples of all basis functions used for model generation. By adding a basis function and its conjugate complex one in every iteration, rFSE generates a real-valued model, even though the basis functions itself are complex-valued. Thus, the prior knowledge of real-valued signals that arises for image and video extrapolation is exploited. As described in detail in \cite{Kaup2005} within every iteration one basis function is selected to be added to the model generated so far and the corresponding expansion coefficient is estimated. In this process, the weighting function
\begin{equation}
\label{eq:weighting_function}
 w\left[m,n\right] = \left\{ \begin{array}{ll} \hat{\rho}^{\sqrt{\left(m-\frac{M-1}{2}\right)^2+\left(n-\frac{N-1}{2}\right)^2}} & \mbox{for } \left(m,n\right)\in \mathcal{A} \\ 0 & \mbox{for } \left(m,n\right)\in \mathcal{B}\end{array}\right.
\end{equation}
is used to control the influence each sample has on the model generation, depending on its position. With that, area $\mathcal{B}$ is masked from the model generation and samples far away from $\mathcal{B}$ obtain only a small weight and therewith low influence on the model generation. The decay of the weighting function in area $\mathcal{A}$ is controlled by $\hat{\rho}$. The iterations of selecting one basis function and estimating its weight are repeated $I$ times. Finally, area $\mathcal{B}$ is cut out of the model and serves for extrapolation of the signal. According to \cite{Kaup2005}, the complete algorithm can be carried out in the Fourier domain and only one transform of the input signal into the frequency domain and one transform of the final model back into the spatial domain are required. For carrying out the transforms efficiently, the Fast Fourier Transform (FFT) \cite{Cooley1965} should be utilized. 

But, although rFSE can operate in the frequency domain more efficiently compared to a spatial domain solution, the individual iterations still possess computationally expensive operations like branches and divisions, as shown in detail in \cite{Kaup2005}. These expensive operations result from the objective of rFSE to generate a real-valued model from complex-valued basis functions. Due to this constraint, the real-valued basis functions with frequency zero and highest possible frequency have to be treated differently than all other basis functions. Furthermore, during the basis function selection in every iteration $M\cdot N$ divisions with variable denominators have to be carried out. Compared to the remaining operations that mainly consist of multiplications and additions the branches and divisions are computationally very expensive and require more time to execute. In order to cope with this, in the next section a complex-valued model generation is introduced which can avoid the just mentioned expensive operations.


\section{Complex-Valued Frequency Selective Extrapolation} \label{sec:cfse} 

In order to reduce the complexity of the model generation, it is advantageous to relax the constraint of generating a real-valued model and generate a complex-valued model instead, even if only real-valued image and video signals are regarded. Due to this, the subsequently regarded algorithm is called complex-valued Frequency Selective Extrapolation (cFSE).  As the limitation of a real-valued model can be omitted, the model generated by cFSE is
\begin{equation}
\label{eq:cFSE_model}
 g\left[m,n\right] = \sum_{\left(k,l\right)\in\mathfrak{K}}\hat{c}_{\left(k,l \right)} \varphi_{\left(k,l \right)}\left[m,n\right] .
\end{equation}
Similar to rFSE, cFSE can be carried out completely in the frequency domain and is the direct translation of the original Selective Extrapolation from \cite{Kaup2005} into the Fourier domain. To achieve this, three properties that emerge from using Fourier basis functions are exploited. Regarding 
\begin{eqnarray}
\label{eq:dft_props_1} \varphi_{\left(k,l \right)}\left[m,n\right] \varphi_{\left(\tilde{k},\tilde{l} \right)}\left[m,n\right] &=& \varphi_{\left(k+\tilde{k},l+\tilde{l}\right)}\left[m,n\right]\\
 \varphi_{\left(k,l \right)}\left[m,n\right] \varphi^\ast_{\left(\tilde{k},\tilde{l} \right)}\left[m,n\right] &=& \varphi_{\left(k-\tilde{k},l-\tilde{l}\right)}\left[m,n\right]
\end{eqnarray}
it becomes obvious that the product of two basis functions is equal to the basis functions resulting from summing up the indices, or respectively, from the difference of the indices if the conjugate complex of one of the basis functions is regarded. If the index resulting from the summation or difference is outside the range between $0$ and $M-1$ or $N-1$, it has to be modulo reduced by $M$ or $N$, respectively.  The third important property is 
\begin{equation}
\label{eq:dft_props_3} \sum_{\left(m,n\right)\in\mathcal{L}} x\left[m,n\right] \varphi^\ast_{\left(k,l \right)}\left[m,n\right] = X\left[k,l\right],
\end{equation}
showing that the summation over the product between the conjugate complex of basis function $\varphi_{\left(k,l \right)}\left[m,n\right]$ and an arbitrary signal $x\left[m,n\right]$, corresponds to the coefficient $X\left[k,l\right]$ of the DFT of $x\left[m,n\right]$ at frequency $\left(k,l\right)$. 

As described in \cite{Kaup2005}, the model from (\ref{eq:cFSE_model}) is generated iteratively, at which in every iteration one basis function is selected and added to the model generated so far. Initially, the model is equal to zero. For generating the model, the approximation residual 
\begin{equation}
r^{\left(\nu-1\right)} \left[m,n\right] = s\left[m,n\right] - g^{\left(\nu-1\right)} \left[m,n\right]
\end{equation}
from the previous iteration is projected onto all basis functions. Here, the weighting function from (\ref{eq:weighting_function}) is used again to control the influence each sample has on the model generation. The weighted projection yields the projection coefficients
\begin{equation}
 p_{\left(k,l \right)}^{\left(\nu\right)} = \frac{\displaystyle \sum_{\left(m,n\right)\in\mathcal{L}} r^{\left(\nu-1\right)} \left[m,n\right] \varphi^\ast_{\left(k,l \right)}\left[m,n\right] w\left[m,n\right]}{\displaystyle \sum_{\left(m,n\right)\in\mathcal{L}} \varphi_{\left(k,l \right)}^\ast\left[m,n\right]w\left[m,n\right]\varphi_{\left(k,l \right)}\left[m,n\right]}. 
\end{equation}
Subsequent to this, the basis function gets selected that minimizes the weighted distance between the residual and the weighted projection onto the corresponding basis function. Hence, the indices of the basis function to select result from
\begin{equation}
 \left(\hspace{-1mm}u^{\left(\nu\right)},v^{\left(\nu\right)}\hspace{-1mm}\right) \hspace{-1mm}=\hspace{-1mm} \argmax_{\left(k,l \right)} \hspace{-1mm}\left(\hspace{-1mm} \left|p_{\left(k,l \right)}^{\left(\nu\right) }\right|^2 \hspace{-0.35cm}\sum_{\left(m,n\right)\in\mathcal{L}}\hspace{-3mm}  \scriptstyle\varphi_{\left(k,l \right)}^\ast\hspace{-0mm}\left[m,n\right]w\hspace{-0mm}\left[m,n\right]\varphi_{\left(k,l \right)}\hspace{-0mm}\left[m,n\right]\hspace{-1mm}\right).
\end{equation}
Using properties (\ref{eq:dft_props_1}) to (\ref{eq:dft_props_3}), the equation above can be evaluated in the frequency domain. For this, the weighted residual
\begin{equation}
 r_w^{\left(\nu-1\right)}\left[m,n\right] =  r^{\left(\nu-1\right)}\left[m,n\right]\cdot w\left[m,n\right]
\end{equation}
 and weighting function $w\left[m,n\right]$ have to be transformed into the frequency domain, yielding  $R_w^{\left(\nu-1\right)}\left[k,l\right]$ and $W\left[k,l\right]$. With that, the calculation of the indices can be eased to
\ifthenelse{\equal{\VERSION}{DRAFT}}{
\begin{equation}
\label{eq:bf_selection_freqdomain}
\left(u^{\left(\nu\right)},v^{\left(\nu\right)}\right) = \argmax_{\left(k,l\right)} \frac{\left|R_w^{\left(\nu-1\right)}\left[k, l\right]\right|^2}{W\left[0,0\right]} = \argmax_{\left(k,l\right)} \left|R_w^{\left(\nu-1\right)}\left[k, l\right]\right|^2
\end{equation}
}{
\begin{equation}
\label{eq:bf_selection_freqdomain}
\left(\hspace{-1mm}u^{\left(\nu\right)},v^{\left(\nu\right)}\hspace{-1mm}\right) \hspace{-1mm}=\hspace{-1mm} \argmax_{\left(k,l\right)} \frac{\left|R_w^{\left(\nu-1\right)}\left[k, l\right]\right|^2}{W\left[0,0\right]} \hspace{-1mm}=\hspace{-1mm} \argmax_{\left(k,l\right)} \left|R_w^{\left(\nu-1\right)}\left[k, l\right]\right|^2
\end{equation}
}
since the denominator is constant and has no influence on the $\argmax$ operation.

For estimating the expansion coefficient $\hat{c}_{\left(u^{\left(\nu\right)},v^{\left(\nu\right)}\right)}$, as well as for rFSE, Fast Orthogonality Deficiency Compensation \cite{Seiler2008} can be applied to cope with the non-orthogonality of the basis function with respect to support area $\mathcal{A}$. With the compensation factor $\gamma$ between $0$ and $1$, depending on the loss scenario, this yields
\begin{equation}
 \hat{c}_{\left(u^{\left(\nu\right)},v^{\left(\nu\right)} \right)} = \gamma \hat{p}_{\left(u^{\left(\nu\right)},v^{\left(\nu\right)} \right)}  = \gamma\frac{R_w^{\left(\nu-1\right)}\left[u^{\left(\nu\right)},v^{\left(\nu\right)}\right]}{W\left[0,0\right]}.
\end{equation}
As the denominator in the equation above is constant, the division can be calculated in advance and be replaced by a multiplication with ${1}/{W\left[0,0\right]}$ within the iteration loop.

After one basis function has been selected and its weight has been estimated, the model has to be updated according to 
\ifthenelse{\equal{\VERSION}{DRAFT}}{
\begin{equation}
  g^{\left(\nu\right)}\left[m,n\right] = g^{\left(\nu-1\right)}\left[m,n\right] + \hat{c}_{\left(u^{\left(\nu\right)},v^{\left(\nu\right)} \right)}\varphi_{\left(u^{\left(\nu\right)},v^{\left(\nu\right)} \right)}\left[m,n\right].
\end{equation}
}{
\begin{equation}
  g^{\left(\nu\right)}\hspace{-0.5mm}\left[m,n\right] \hspace{-1mm}=\hspace{-1mm} g^{\left(\nu-1\right)}\hspace{-0.5mm}\left[m,n\right] \hspace{-1mm}+\hspace{-1mm} \hat{c}_{\left(u^{\left(\nu\right)},v^{\left(\nu\right)} \right)}\varphi_{\left(u^{\left(\nu\right)},v^{\left(\nu\right)} \right)}\left[m,n\right].
\end{equation}
}
With $G^{\left(\nu\right)}\left[k,l\right]$ being the DFT of $g^{\left(\nu\right)}\left[m,n\right]$, the update can as well be carried out in the Fourier domain by 
\ifthenelse{\equal{\VERSION}{DRAFT}}{
\begin{equation}
\label{eq:model_update_freqdomain}
G^{\left(\nu\right)}\left[u^{\left(\nu\right)},v^{\left(\nu\right)}\right] = G^{\left(\nu-1\right)}\left[u^{\left(\nu\right)},v^{\left(\nu\right)}\right] + MN\hat{c}_{\left(u^{\left(\nu\right)},v^{\left(\nu\right)} \right)}.
\end{equation}
}{
\begin{equation} 
\label{eq:model_update_freqdomain}
G^{\left(\nu\right)}\hspace{-1mm}\left[u^{\left(\nu\right)},v^{\left(\nu\right)}\right] \hspace{-1mm}=\hspace{-1mm} G^{\left(\nu-1\right)}\hspace{-1mm}\left[u^{\left(\nu\right)},v^{\left(\nu\right)}\right] \hspace{-1mm}+\hspace{-1mm} MN\hat{c}_{\left(u^{\left(\nu\right)},v^{\left(\nu\right)} \right)}.
\end{equation}
}
Apparently, only the coefficient of the transformed model that belongs to the selected basis function has to be modified, as
\ifthenelse{\equal{\VERSION}{DRAFT}}{
\begin{equation}
 \sum_{\left(m,n\right)\in\mathcal{L}}\varphi_{\left(u^{\left(\nu\right)},v^{\left(\nu\right)}\right)}\varphi^\ast_{\left(k,l\right)} = \left\{\begin{array}{ll} MN & \mbox{for }\left(u^{\left(\nu\right)},v^{\left(\nu\right)}\right) = \left(k,l\right)\\0 &\mbox{else}\end{array}\right.
\end{equation}
}{
\begin{equation}
 \sum_{\left(m,n\right)\in\mathcal{L}}\hspace{-2mm}\varphi_{\left(u^{\left(\nu\right)},v^{\left(\nu\right)}\right)}\varphi^\ast_{\left(k,l\right)} = \left\{\hspace{-1mm}\begin{array}{ll} MN & \hspace{-1mm}\mbox{for }\left(u^{\left(\nu\right)},v^{\left(\nu\right)}\right) = \left(k,l\right)\\0 &\hspace{-1mm}\mbox{else}\end{array}\right.
\end{equation}
}
holds for the scalar product between two basis functions. Since the DFT of the weighted residual is required in (\ref{eq:bf_selection_freqdomain}), the residual update 
\ifthenelse{\equal{\VERSION}{DRAFT}}{
\begin{equation}
r^{\left(\nu\right)}\left[m,n\right] = r^{\left(\nu-1\right)}\left[m,n\right] - \hat{c}_{\left(u^{\left(\nu\right)},v^{\left(\nu\right)} \right)} \varphi_{\left(u^{\left(\nu\right)},v^{\left(\nu\right)} \right)} \left[m,n\right]
\end{equation}
}{
\begin{equation}
r^{\left(\nu\right)}\hspace{-0.5mm}\left[m,n\right] \hspace{-1mm}=\hspace{-1mm} r^{\left(\nu-1\right)}\hspace{-0.5mm}\left[m,n\right] \hspace{-1mm}-\hspace{-1mm} \hat{c}_{\left(u^{\left(\nu\right)},v^{\left(\nu\right)} \right)} \varphi_{\left(u^{\left(\nu\right)},v^{\left(\nu\right)} \right)} \left[m,n\right]
\end{equation}
}
is not transformed directly into the Fourier domain. Instead, the update of the weighted residual $r_w^{\left(\nu\right)}\left[m,n\right]$ is regarded which can be expressed in the Fourier domain by
\ifthenelse{\equal{\VERSION}{DRAFT}}{
\begin{equation}
 R_w^{\left(\nu\right)} \left[k,l\right] = R_w^{\left(\nu-1\right)} \left[k,l\right] - \hat{c}_{\left(u^{\left(\nu\right)},v^{\left(\nu\right)} \right)} W\left[k-u^{\left(\nu\right)}, l-v^{\left(\nu\right)}\right], \forall\left(k,l\right).
\end{equation}
}{
\begin{equation}
 R_w^{\left(\nu\right)} \hspace{-1mm}\left[k,l\right] \hspace{-1mm}=\hspace{-1mm} R_w^{\left(\nu-1\right)} \hspace{-1mm}\left[k,l\right] - \hat{c}_{\left(u^{\left(\nu\right)},v^{\left(\nu\right)} \right)} \hspace{-0.5mm}W\hspace{-1mm}\left[\hspace{-0.5mm}k\hspace{-1mm}-\hspace{-1mm}u^{\left(\nu\right)}, l\hspace{-1mm}-\hspace{-1mm}v^{\left(\nu\right)}\hspace{-0.5mm}\right]\hspace{-0.5mm},\hspace{-0.5mm} \forall\hspace{-0.5mm}\left(k,l\right).
\end{equation}
}
Here, unlike (\ref{eq:model_update_freqdomain}), all frequency bins have to be updated as the weighting function is included in the scalar product calculation.

These steps are repeated $I$ times. According to the derivation shown above, the algorithm can also be carried out completely in the Fourier domain. Thus, only one transform into the Fourier domain prior to the model generation and one back into the spatial domain after the iterations have been finished are required. Finally, the model in area $\mathcal{B}$ serves for extrapolation of signal $s\left[m,n\right]$.

To give a compact overview, Alg.\ \ref{algo:fse_cbf} shows the pseudo code of cFSE. It can be recognized, that the iteration loop of cFSE does not contain any branches or divisions anymore. Hence, the model generation for cFSE is less complex than the one of rFSE and can be carried out very fast. If the signal to be extrapolated is known to be real-valued, this knowledge can be exploited by discarding the complex-valued part of the generated model at the end. 

\begin{algorithm}
\ifthenelse{\equal{\VERSION}{DRAFT}}{\renewcommand{\baselinestretch}{1}}{}
\caption{Complex-valued Frequency Selective Extrapolation}
\small
\label{algo:fse_cbf}
\begin{algorithmic}
\INPUT distorted signal $s\left[m,n\right]$, weighting function $w\left[m,n\right]$
	\STATE /* Transform input signals into Fourier domain */
	\STATE $R_w\left[k,l\right] = \mathrm{FFT}\left\{s\left[m,n\right]w\left[m,n\right]\right\}$
	\STATE $W\left[k,l\right] = \mathrm{FFT}\left\{w\left[m,n\right]\right\}$
	\STATE $\bar{W}_0 = \frac{1}{W\left[0,0\right]}$    
	\FORALL {$\nu=1, \ldots, I$}
		\STATE /* Basis function selection */
		\STATE $\left(u,v\right) = \argmax_{\left(k,l\right)} \left|R_w\left[k,l\right]\right|^2$
		\STATE /* Expansion coefficient estimation */
		\STATE $\hat{c} = \gamma R_w\left[u,v\right]\bar{W}_0$
		\STATE /* Model update */
		\STATE $G\left[u,v\right] = G\left[u,v\right] + MN\hat{c}$
		\STATE /* Residual update */
		\FORALL {$k=0,\ldots,M-1 \wedge l=0,\ldots,N-1$}
			\STATE $R_w\left[k,l\right] = R_w\left[k,l\right] -\hat{c} W\left[k-u, l-v\right]$
		\ENDFOR
	\ENDFOR
	\STATE /* Retransform model into spatial domain*/
	\STATE $g\left[m,n\right] = \mathrm{IFFT}\left\{G\left[k,l\right]\right\}$
	\STATE /* Replace distorted signal parts */
	\FORALL {$\left(m,n\right)\in\mathcal{B}$}
		\STATE $s\left[m,n\right] = \mathrm{Re}\left\{g\left[m,n\right]\right\}$
	\ENDFOR
\OUTPUT extrapolated signal $s\left[m,n\right]$
\end{algorithmic}
\end{algorithm}


\section{Simulations and Results}\label{sec:results} 

For evaluating the computational complexity of cFSE compared to rFSE, both algorithms are tested for concealment of isolated block losses. For this purpose, the algorithms have been implemented in C, compiled with gcc4.3 and run on an Intel Pentium D@$3.20 \punit{GHz}$, equipped with $6 \punit{GB}$ RAM. The losses to be concealed are of size $16\times 16$ samples, the support area is $16$ samples wide, the weighting function declines with $\hat{\rho}=0.8$, the orthogonality deficiency compensation is set to $\gamma=0.2$ and an FFT of size $64\times 64$ is used. For the transform into the Fourier domain the FFTW3 is utilized. 

\begin{figure}
	\centering
	\psfrag{a}{a)}
	\psfrag{b}{b)}
	\psfrag{c}{c)}
	\includegraphics[draft=FALSE,width=0.4\textwidth]{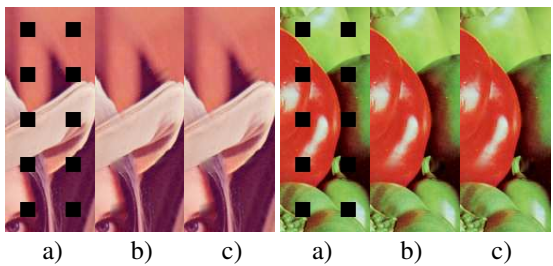}
	\caption{Visual results for concealment of isolated blocks of size $16\times 16$ samples. a) Error pattern, b) Extrapolation by rFSE, c) Extrapolation by cFSE.}\vspace{-0.5cm}
	\label{fig:visual_results}
\end{figure}

\begin{figure}
	\centering
	\input{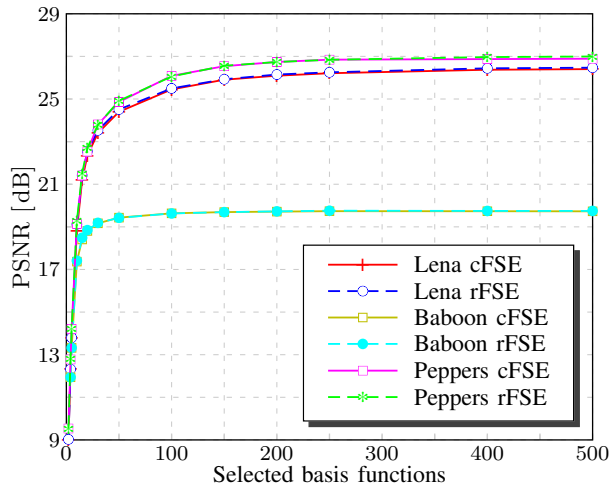}
	\caption{Extrapolation quality over number of selected basis functions for concealment of isolated block losses of size $16\times 16$ samples.}
	\label{fig:quality}
\end{figure}

Since rFSE directly exploits the knowledge of the real-valued input signal for model generation, it should be able to achieve a higher extrapolation quality than cFSE. To quantify this, Fig.\ \ref{fig:quality} shows the extrapolation quality in $\PSNR$ over the number of basis functions used for model generation. Multiple selections of the same basis functions are counted individually. As rFSE selects one basis function and its conjugate complex in every iteration, except for the constant and highest alternating frequency, rFSE only has to perform roughly half as many iterations as cFSE for selecting the same number of basis functions. Comparing the curves for rFSE and cFSE, no significant discrepancy can be discovered for any of the test sequences. Regarding all test images from the Kodak test image data base, the mean gain of rFSE over cFSE is only $0.035 \punit{dB}$. In addition to that, Fig.\ \ref{fig:visual_results} shows visual examples for concealment of block losses by rFSE and cFSE. As well as for the objective $\PSNR$ evaluation, no difference can be discovered and rFSE and cFSE are both able to achieve a very high visual extrapolation quality. Taking these results into account, it can be discovered that the extrapolation quality of rFSE is only negligibly superior to the one of cFSE.

\begin{figure}
	\centering
	\input{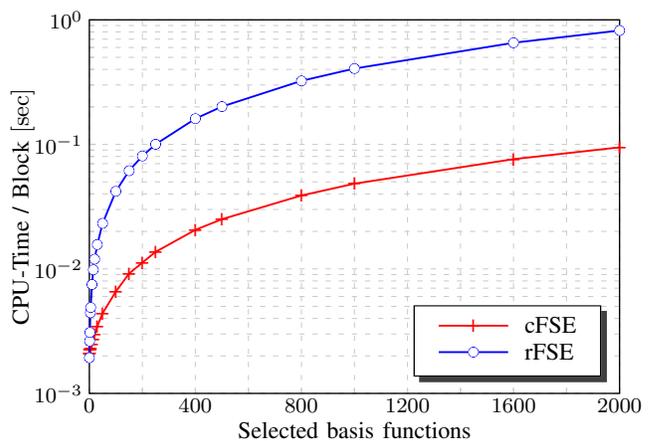}
	\caption{Processing time over number of selected basis functions for 2D model generation with DFT basis functions of size $64\times 64$.}
	\label{fig:runtime}
\end{figure}

Fig.\ \ref{fig:runtime} shows the extrapolation time per block for carrying out the extrapolation by rFSE and cFSE with different numbers of selected basis functions. Comparing the two curves, one can discover, that cFSE is able to outspeed rFSE significantly, resulting in a factor of up to $10$ for large numbers of selected basis functions. For small numbers of selected basis functions, the influence of the initial FFT and final inverse FFT is recognizable, but this influence diminishes with an increasing number of selected basis functions.
Thus, the plain iteration loop of cFSE without divisions and branches takes effect and contributes to the overall accelerated model generation.

\section{Conclusion} \label{sec:conclusion} 
 
In this contribution the complex-valued Frequency Selective Extrapolation was introduced for image and video signal extrapolation. This algorithm iteratively generates a model of the signal to be extrapolated as weighted superposition of Fourier basis functions. Compared to the original real-valued Frequency Selective Extrapolation an acceleration by a factor of up to $10$ is possible since the model generation does not require branches or divisions. At the same time, nearly the same extrapolation quality can be achieved with almost no loss in $\PSNR$.

For presentational reasons, the algorithm has been introduced only for two-dimensional signal extrapolation, but, as shown in \cite{Meisinger2007}, it can be easily extended to higher dimensional problems as well. For higher dimensional data sets, the proposed complex-valued Frequency Selective Extrapolation becomes even more important, as the original real-valued extrapolation suffers from the increasing number of branches and division that arise with higher dimensions.

The avoidance of divisions furthermore is beneficial for fixed-point implementations. As divisions can be carried out with a reduced accuracy only in this context, the repeated use of divisions can lead to error propagation. Since the proposed complex-valued model generation can avoid all divisions, it eliminates this risk.

\centerline{\rule{0.45\textwidth}{0.4pt}}

\ifthenelse{\equal{\VERSION}{DRAFT}}{\renewcommand{\baselinestretch}{1}}{\renewcommand{\baselinestretch}{0.8}}


\end{document}